\documentclass[11pt]{article}

\usepackage[a4paper,margin=1in]{geometry}
\usepackage{authblk}
\usepackage{amsmath,amssymb,amsfonts}
\usepackage{graphicx}
\usepackage{multirow}
\usepackage{booktabs}
\usepackage{xcolor}
\usepackage{algorithm}
\usepackage{algpseudocode}
\usepackage[numbers,sort&compress]{natbib}
\usepackage{newunicodechar}
\usepackage{hyperref}

\newunicodechar{ℓ}{\ensuremath{\ell}}

\hypersetup{
    colorlinks=true,
    linkcolor=blue,
    citecolor=blue,
    urlcolor=blue
}

\newcommand{\purple}[1]{#1}

\let\bmsection\section
\let\bmsubsection\subsection
\let\bmsubsubsection\subsubsection

\providecommand{\authormark}[1]{}
\providecommand{\titlemark}[1]{}

\providecommand{\orgdiv}[1]{#1}
\providecommand{\orgname}[1]{#1}
\providecommand{\orgaddress}[1]{#1}

\providecommand{\city}[1]{#1}
\providecommand{\postcode}[1]{#1}

\providecommand{\country}[1]{#1}

\providecommand{\address}{}
\renewcommand{\address}[2][1]{\affil[#1]{#2}}

\newcommand{\email}[1]{\space\textless\href{mailto:#1}{\texttt{#1}}\textgreater}

\makeatletter
\newcommand{\@CorresInfo}{}
\newcommand{\@AbstractTitle}{Abstract}
\newcommand{\@AbstractBody}{}
\newcommand{\@KeywordsBody}{}
\providecommand{\corres}[1]{}
\renewcommand{\corres}[1]{\gdef\@CorresInfo{#1}}
\renewcommand{\abstract}[2][Abstract]{\gdef\@AbstractTitle{#1}\gdef\@AbstractBody{#2}}
\providecommand{\keywords}[1]{}
\renewcommand{\keywords}[1]{\gdef\@KeywordsBody{#1}}

\let\OrigMaketitle\maketitle
\renewcommand{\maketitle}{%
  \OrigMaketitle
  \ifx\@CorresInfo\@empty\else
    {\par\noindent\textit{Correspondence:}\space\@CorresInfo\par\medskip}%
  \fi
  \ifx\@AbstractBody\@empty\else
    \begin{center}\bfseries\large\@AbstractTitle\end{center}%
    \begin{quote}\noindent\@AbstractBody\par\end{quote}\medskip
  \fi
  \ifx\@KeywordsBody\@empty\else
    \par\noindent\textbf{Keywords:}\space\@KeywordsBody\par\medskip
  \fi
}
\makeatother

\begin{document}

\newcommand{\Title}{NexOP: Joint Optimization of NEX-Aware \textit{$k$-space} Sampling and Image Reconstruction for Low-Field MRI}

\title{\Title}

\author[1,2]{Tal Oved}

\author[1,2,3]{Efrat Shimron}

\authormark{Oved \textsc{et al.}}
\titlemark{NexOP}  


\address[1]{%
  \orgdiv{Department of Electrical and Computer Engineering}, 
  \orgname{Technion – Israel Institute of Technology}, 
  \orgaddress{%
    \city{Haifa}, 
    \postcode{3200004}, 
    \country{Israel}}}

 \address[2]{\orgdiv{May-Blum-Dahl Technion Human MRI Research Center}, \orgname{Technion - Israel Institute of Technology}, \orgaddress{ 
\city{Haifa}, \postcode{3200004}, 
 \country{Israel}}}

\address[3]{%
  \orgdiv{Department of Biomedical Engineering}, 
  \orgname{Technion – Israel Institute of Technology}, 
  \orgaddress{%
    \city{Haifa}, 
    \postcode{3200004}, 
    \country{Israel}}}


\corres{Corresponding author Efrat Shimron\email{efrat.s@technion.co.il}}

\abstract[Abstract]{Modern low-field magnetic resonance imaging (MRI) technology offers a compelling alternative to standard high-field MRI, with portable, low-cost systems. However, its clinical utility is limited by a low Signal-to-Noise Ratio (SNR), which hampers diagnostic image quality. A common approach to increase SNR is through repetitive signal acquisitions, known as NEX, but this results in excessively long scan durations. Although recent work has introduced methods to accelerate MRI scans through \textit{$k$-space} sampling optimization, the NEX dimension remains unexploited; typically, a single sampling mask is used across all repetitions. Here we introduce NexOP, a deep-learning framework for joint optimization of the sampling and reconstruction in multi-NEX acquisitions, tailored for low-SNR settings. NexOP enables optimizing the sampling density probabilities across the extended \textit{$k$-space}-NEX domain, under a fixed sampling-budget constraint, and introduces a new deep-learning architecture for reconstructing a single high-SNR image from multiple low-SNR measurements. Experiments with raw low-field ($0.3\,\mathrm{T}$) brain data demonstrate that NexOP consistently outperforms competing methods, both quantitatively and qualitatively, across diverse acceleration factors and tissue contrasts. The results also demonstrate that NexOP yields non-uniform sampling strategies, with progressively decreasing sampling across repetitions, hence exploiting the NEX dimension efficiently. Moreover, we present a theoretical analysis supporting these numerical observations.   Overall, this work proposes a sampling-reconstruction optimization framework highly suitable for low-field MRI, which can enable faster, higher-quality imaging with low-cost systems and contribute to advancing affordable and accessible healthcare.   
} 

\keywords{MRI, Magnetic Resonance Imaging, Optimization, Sampling, Deep Learning, Machine Learning, NEX, Acceleration}

\maketitle

\renewcommand\thefootnote{}
\footnotetext{\textbf{Abbreviations:} SNR, signal-to-noise-ratio; NEX, number-of-excitations; NA, number-of-averages; VD, variable-density; DC, data consistency; ACS, auto-calibration signal; MR step, multi-repetition-step.}

\renewcommand\thefootnote{\fnsymbol{footnote}}
\setcounter{footnote}{1}

\section{Introduction}\label{sec1}

Low-field magnetic resonance imaging (MRI) is rapidly emerging as a transformative technology for accessible and point-of-care imaging \cite{geethanath2019accessible, sarracanie2015low,marques2019low, campbell2023low, Kimberly2023, guallart2022portable,zhao2024whole, ayde2025mri,shimron2024accelerating,oved2025deep}. Compared to conventional high-field ($1.5-3\,\mathrm{T}$) systems, low-field scanners (typically $\sim$$0.005 -0.5\,\mathrm{T}$) offer substantially lower cost, portability, reduced infrastructure demands, and improved safety profiles, hence they enable deployment in new settings, including resource-limited regions, outpatient centers, and at the patient's bedside \cite{geethanath2019accessible,marques2019low}. However, because the signal-to-noise ratio (SNR) scales with magnetic field strength \cite{hoult1976signal}, low-field MRI operates in SNR-starved regimes and suffer from reduced image quality compared to high-field alternatives \cite{marques2019low,campbell2023low,ayde2025mri}.

To compensate for the inherent SNR deficit at low field, imaging protocols often incorporate multiple repeated excitations followed by \textit{$k$-space} data averaging. The number of repetitions is commonly termed the \textit{number of excitations} (NEX) or \textit{number of averages} (NA). Although increasing the NEX improves the SNR, it also increases scan duration in a linear manner, creating a critical quality-versus-time trade-off. Furthermore, as low-field systems frequently rely on time-consuming volumetric 3D acquisitions to achieve sufficient signal from tissue, the addition of multiple repetitions can result in prohibitively long durations, with studies reporting  30-60 minute scans \cite{espy2014progress, shimron2024accelerating}. Therefore, there is a pressing need for advanced acceleration techniques that can reduce low-field MRI scan duration without compromising diagnostic utility \cite{campbell2023low}.


Over the last few decades, a large body of work has focused on developing methods to reduce MRI scan duration, often utilizing sub-Nyquist \textit{$k$-space} sampling and computational techniques that solve the associated ill-posed image-reconstruction problem. Parallel Imaging \cite{griswold2002generalized,pruessmann1999sense,lustig2010spirit,shimron2019core} and Compressed sensing (CS) \cite{Lustig2007,geethanath2013compressed} are two well-established frameworks developed for this aim. Recently, deep learning (DL)-based frameworks have demonstrated state-of-the-art performance, enabling superb-quality reconstructions from highly undersampled data \cite{Heckel2024,hammernik2023physics,Aggarwal2023}. Recent work has thus begun to explore the development of DL-based techniques for low-field MRI reconstruction \cite{zhou2022dual,man2023deep, Koonjoo2021,ilicak2025physics,shimron2023ai,oved2025deep}.

In parallel, the development of methods for optimizing  \textit{$k$-space} sampling strategies has also become an active research area  \cite{bahadir2020deep,chaithya2022optimizing,Aggarwal2020,Weiss2021,ravula2023optimizing,radhakrishna2023jointly,  Gautam2025, Alkan2020}. These efforts have focused on optimizing sampling densities \cite{knoll2011adapted,chauffert2014variable,chaithya2021learning}, sampling masks defined over Cartesian grids \cite{bahadir2020deep,ravula2023optimizing}, and non-Cartesian sampling trajectories defined over a continuous \textit{$k$-space} \cite{Weiss2021,chaithya2022optimizing}. Additionally, recent studies explored joint sampling-reconstruction optimization paradigms, to further improve reconstruction quality \cite{bahadir2020deep,Aggarwal2020,radhakrishna2023jointly}.

However, these research efforts have primarily focused on optimizing sampling for a \textit{single} \textit{$k$-space} acquisition, without leveraging the additional flexibility offered by repeated acquisitions in low-field MRI. Yet, recently, two studies have shown that for a fixed scan-duration budget, combining accelerated sampling with repeated acquisitions can improve reconstruction quality \cite{Schoormans2020,shimron2023ai}. However, these studies used an \textit{identical, fixed} sampling mask across all repetitions; they have not explored modern DL-based sampling optimization techniques. There is hence unexploited potential for more effective use of the repetition dimension in low-SNR imaging.

Here, we introduce NexOP, a NEX-aware framework for jointly optimizing \textit{$k$-space} sampling and image reconstruction, tailored for low-SNR settings, with a focus on low-field MRI. Unlike prior approaches that optimize sampling masks for a single acquisition, NexOP operates over the joint \textit{$k$-space}-NEX domain, learning both where to sample and how to allocate samples across repetitions under a fixed acquisition budget. Moreover, we introduce a new DL architecture that leverages shared information across repetitions to reconstruct a single high-SNR image from multiple low-SNR measurements.

The main contributions of this work are:
\begin{enumerate}
     \item Proposing NexOP, an end-to-end framework for joint optimization of the sampling and reconstruction in low-SNR,  multi-NEX acquisitions, typical of low-field MRI. Our framework enables efficient distribution of \textit{$k$-space} samples across the NEX dimension, which was seldom leveraged in prior methods.
    \item Introducing a new DL reconstruction architecture that reconstructs a single high-SNR image from multiple undersampled, low-SNR acquisitions by leveraging shared features across this dimension, along with a learned deep prior.
    \item Demonstrating the proposed framework on \textit{raw} $0.3\,\mathrm{T}$ MRI data, across T1-weighted and T2-weighted contrasts and a range of acceleration factors. In contrast to the common approach of repeating a single sampling mask, the results suggest that the optimal sampling strategy involves wider \textit{$k$-space} sampling in the first acquisition and progressively decreasing sampling with subsequent repetitions. \purple{Moreover, the results indicate that the optimal sampling strategy exhibits varying \textit{acceleration factors} across the repetitions.}
    \item Releasing open code to support reproducibility and further development of joint sampling-reconstruction frameworks for low-field MRI.
\end{enumerate}

\begin{figure*}
\centerline{\includegraphics[width=1\textwidth]{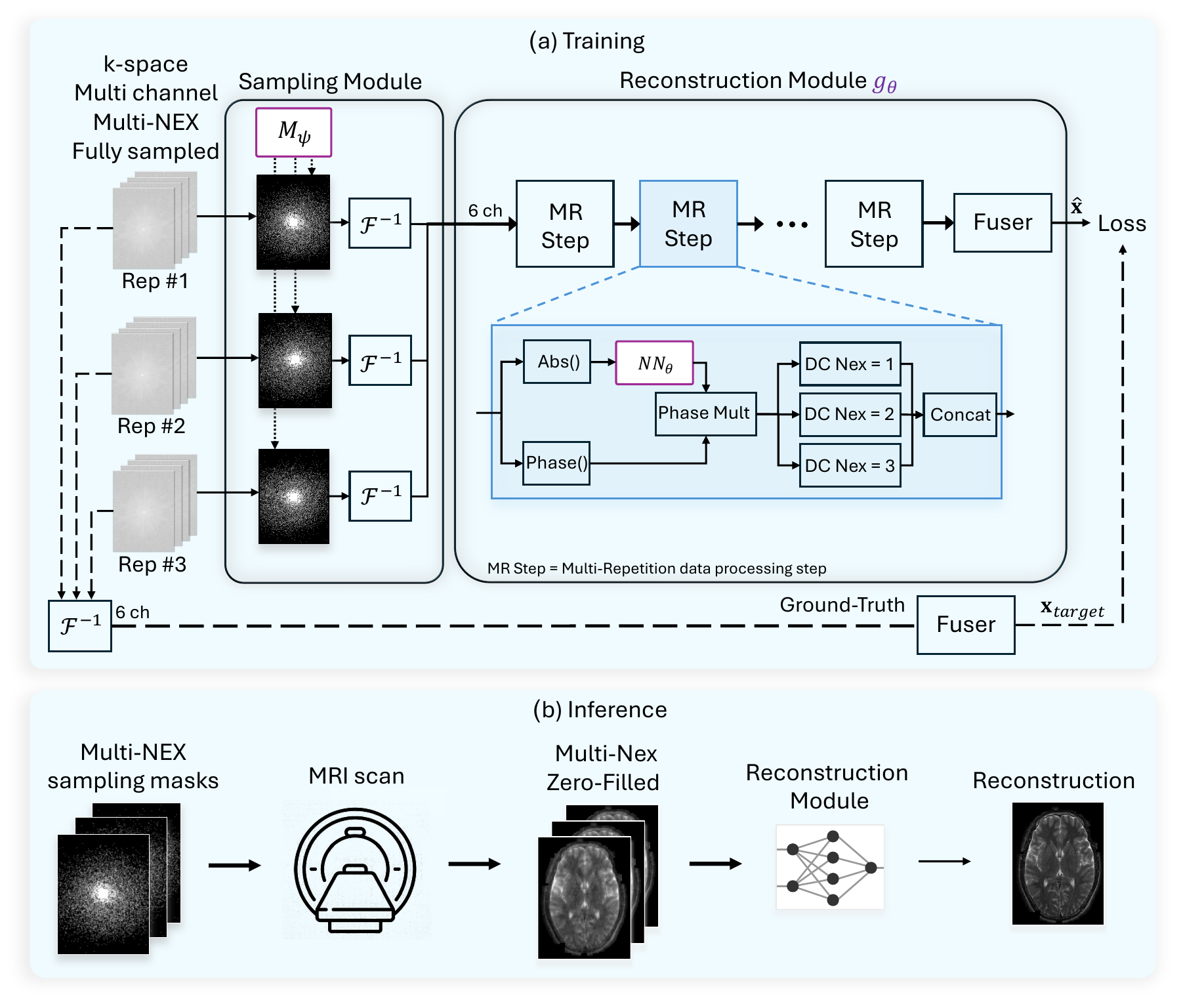}}
\caption{\textbf{Overview of the NexOP framework.} NexOP is a framework designed for joint optimization of the \textit{$k$-space} sampling and reconstruction in multi-NEX MRI under a fixed acquisition budget.
(a) Training pipeline: The \textit{Sampling Module} (left) optimizes a set of learnable parameters $\psi$ that define the multi-repetition (NEX) sampling masks. During training, the 3-repetition undersampled \textit{$k$-space} are inverse-Fourier-transformed, separated into real and imaginary parts, and concatenated, resulting in a 6-channel data stream, which is fed to the \textit{Reconstruction Module} $g_{\theta}$ (right). This module comprises cascaded Multi-Repetition Steps (MR), which separates the signal into magnitude and phase; the magnitude part is processed via a neural network ($NN_\theta$), which exploits shared multi-NEX information to remove aliasing artifacts, while the original phase is multiplied back for Data Consistency (DC) blocks. Eventually, the data are passed through a Fuser block, which produces a single high-SNR image.
(b) Inference: The optimized parameters are fixed and used to generate sampling masks for data acquisition, after which the data are zero-filled and passed through the trained reconstruction network to produce the final output. \label{fig1}
} 
\end{figure*}

\vspace{-1.5em}
 
\section{Materials and Methods}\label{sec2}

\bmsubsection{\textbf{Reconstruction problem}}

In MRI, measurements are inherently acquired in the Fourier domain, which is known as \textit{$k$-space}. This process is often formulated by:
\begin{equation} 
\mathbf{y} = MA\mathbf{x} + \boldsymbol{\epsilon}
\label{eq1_y_from_x}
\end{equation} 
where $\mathbf{x} \in \mathbb{C}^{D}$ is the underlying magnetic resonance image (i.e, transverse magnetization image),
$A$ is a forward operator comprising the Fourier transform, denoted by $F$, $M$ is a binary sampling operator that equals one for the acquired \textit{$k$-space} locations and zero otherwise, $\mathbf{y} \in \mathbb{C}^{D}$ is a vector of acquired \textit{$k$-space} measurements, represented on a fully sampled (i.e., Nyquist-rate sampled) \textit{$k$-space} grid with zeros at unsampled locations, and $\boldsymbol{\epsilon}$ is the measurement noise, which is commonly assumed to be additive white Gaussian noise, i.e.,   $\boldsymbol{\epsilon} \sim \mathcal{N}(0, \sigma^2 \mathbf{I})$, where $\mathbf{I}$ is the identity matrix. \purple{This formulation assumes a single acquisition per sampled \textit{$k$-space} location, resulting in independent and identically distributed noise with a uniform variance of $\sigma^2$.} In case of a multi-coil acquisition, the forward operator $A$ can also include the sensitivity weighting operator $S$, i.e., $A = FS$.   

This work assumes \textit{$k$-space} sampling on a Cartesian grid, which is the most common sampling scheme used in clinical MRI 
\cite{bernstein2004handbook, deshmane2012parallel}. Due to practical acquisition considerations, the MRI readout dimension is always fully sampled, and undersampling (i.e., sub-Nyquist sampling) is performed primarily along the phase-encoding dimension(s) \cite{bernstein2004handbook,hollingsworth2015reducing,deshmane2012parallel, Lustig2007}. 
Consequently, in 3D acquisitions, a 2D undersampling mask is commonly applied to the two phase-encoding dimensions. Common undersampling strategies include equi-spaced, random-uniform, and random variable-density (VD) patterns \cite{tsai2000reduced, Lustig2007, chauffert2013variable}. VD schemes sample the \textit{$k$-space} center more densely than the periphery and produce incoherent, noise-like artifacts that can be effectively removed by iterative reconstruction algorithms, hence they are commonly used \cite{Lustig2007,Heckel2024}.
 
When undersampling is applied, the MRI reconstruction task becomes an ill-posed inverse problem, and is commonly addressed through computational methods that utilize inherent data redundancy or leverage \textit{a-priori} knowledge about the image, such as transform-domain sparsity. Well-established examples include parallel imaging and compressed sensing methods \cite{pruessmann1999sense,Lustig2007, lustig2010spirit,shimron2019core,geethanath2013compressed,deshmane2012parallel, zhu2018image,Heckel2024}.

Recently, modern DL strategies have achieved state-of-the-art performance in solving MRI inverse problems, using a variety of architectures and algorithmic approaches \cite{Hammernik2017LearningAV, aggarwal2018modl, Heckel2024, hammernik2023physics, Sharma2022, shimron2020temporal, Gu2022, Korkmaz2021, Kabas2024, Desai2023, shimron2023ai,wang2023k}. DL-based frameworks commonly address the reconstruction task by employing a parameterized neural network, $f_\theta$, to map an initial aliased image, computed by zero-filling (ZF) the undersampled \textit{$k$-space} data, into an alias-free reconstructed image $\hat{\mathbf{x}}$.  This is typically done using supervised training with a dataset consisting of $N$ examples $\{\mathbf{x}_\text{gt}^{(i)}, \mathbf{y}^{(i)}\}_{i=1}^N$, where $\mathbf{x}_\text{gt}^{(i)}$  are ground-truth images corresponding to fully sampled \textit{$k$-space} data $\mathbf{y}_\text{gt}^{(i)}$, and $\mathbf{y}^{(i)}$ are their undersampled counterparts, as in Eq.  \eqref{eq1_y_from_x}. The parameters of $f_{\theta}$ are commonly optimized by minimizing a loss function  such as
\begin{equation}
\mathcal{L}(\theta) = \frac{1}{N} \sum_{i=1}^{N} \left\|   f_{\theta}(A^{H}\mathbf{y}^{(i)})-\mathbf{x}_\text{gt}^{(i)} \right\|_2^2
\label{eq3_ClassicLoss} 
\end{equation}
where $A^{H}$ is the adjoint of the fully sampled forward-process operator $A$; here $H$ denotes the Hermitian transpose. In the multi-coil case, $A^{H}$  also includes a sensitivity weighted coil-combination \cite{roemer1990nmr}. The term $A^H \mathbf{y}^{(i)}$ (which equals $A^H M A \mathbf{x}^{(i)}$ in a noiseless setting) represents the initial aliased image in the spatial domain, which serves as the input to the reconstruction network.
While Eq.~\eqref{eq3_ClassicLoss} employs an $\ell_2$ loss, other loss functions are also common, e.g., $\ell_1$ and feature-based losses \cite{Heckel2024}.

\bmsubsection{\textbf{Joint sampling-reconstruction optimization for a \textit{single} acquisition}}\label{sec_LOUPE}

Before introducing our NexOP framework, we describe LOUPE, a well-established framework for jointly optimizing sampling and reconstruction in MRI \cite{bahadir2020deep}, which serves as a baseline in this study. We note that it was developed for a \textit{single} ($NEX=1$) \textit{$k$-space} acquisition.   

\textbf{\textit{Sampling budget definition.}} In the LOUPE framework, the acquisition constraint is parameterized by a target \emph{sparsity} $\alpha \in (0,1)$, defined as the fraction of \textit{$k$-space} locations to be acquired; the acceleration is therefore $R = 1/\alpha$. For consistency with the multi-NEX formulation introduced in later sections, we define an equivalent \emph{sampling budget} $B := \alpha\,D$, where $D$ is the total number of discrete locations on the fully sampled \textit{$k$-space} grid, and $B$ is the total number of acquired \textit{$k$-space} locations out of $D$, i.e., $B<D$. Therefore, $\alpha = B/D = 1/R$. Given this budget, or equivalently this sparsity, the goal of joint sampling-reconstruction optimization is to learn both a reconstruction network and a budget-feasible sampling distribution (i.e.\ how to allocate the $B$ samples across \textit{$k$-space} locations) so as to minimize the reconstruction error.

\bmsubsubsection*{LOUPE framework}

In the LOUPE framework, the fixed sampling operator $M$ is replaced by a learnable binary sampling operator $M_\phi \in \{0,1\}^{D}$, where $\phi \in \mathbb{R}^D$ denotes a vector of unconstrained real-valued parameters, also known as \textit{logits}. The acquisition process is then modeled by:
\begin{equation}
\mathbf{y}_{\phi} = M_{\phi} A\mathbf{x} + \boldsymbol{\epsilon} 
\label{eq:learned_no_nex_operator}
\end{equation}
where $\mathbf{y}_{\phi}$ are the undersampled measurements. A fundamental challenge in optimizing $M_{\phi}$ is its discrete nature; because it consists of binary values, the mapping from the learnable parameters $\phi$ to the sampling mask is non-differentiable, which prevents standard gradient-based backpropagation.

To address this, the discrete nature of the mask is relaxed by introducing a probabilistic sampling map $\mathbf{P} \in (0,1)^D$. In this map, each entry $p_k$ represents the probability that the $k$-th location of \textit{$k$-space} will be selected, effectively serving as the parameter for a Bernoulli distribution at that location. However, directly optimizing parameters constrained to a bounded interval is problematic for standard gradient-based optimizers, which operate most effectively on unconstrained values. To resolve this, $\mathbf{P}$ is parameterized using the unconstrained real-valued vector $\phi \in \mathbb{R}^D$ introduced above, such that $p_k = \sigma(\phi_k)$, where $\sigma(a)=\frac{1}{1+e^{-a}}$ is the standard sigmoid function. This ensures that for any real-valued logit $\phi_k \in (-\infty, \infty)$, the resulting sampling probability $p_k$ is mapped to the valid probability range $(0,1)$. This formulation allows the framework to optimize the sampling pattern in a continuous, unconstrained space, before generating the final binary mask $\mathbf{M}_\phi \in \{0,1\}^D$ during the forward pass.

Next, we describe how the sampling-budget constraint is enforced during the optimization process, first as in the original LOUPE framework via a sigmoid-based relaxation \cite{bahadir2020deep}, and then via the Gumbel-Softmax implementation used in this work.

\bmsubsubsection*{LOUPE's sigmoid-based sampling optimization}

In the original LOUPE framework \cite{bahadir2020deep}, the sigmoid used to map the logits $\phi_k$ to probabilities $p_k$ is a slope-parameterized variant $\sigma_t(\cdot)$, where the slope $t > 0$ is a hyperparameter that controls its sharpness. The budget is then enforced by a dedicated piecewise normalization layer $\mathcal{N}_\alpha$ that is applied to the probability vector $\mathbf{p}$, producing $\mathbf{q}:=\mathcal{N}_\alpha(\mathbf{p})$ with entries $q_k$ given element-wise by Eq.~5 in Bahadir et al.\ \cite{bahadir2020deep}, whose coefficient depends on the global average $\bar{p}=\tfrac{1}{D}\sum_{j} p_j$. The layer is designed so that $q_k \in [0,1]$ adheres to the target sparsity. The realized binary mask remains stochastic. Finally, to enable end-to-end differentiability, the non-differentiable binary draw is approximated by a uniform-noise, sigmoid-smoothed thresholding applied to the normalized probability $q_k$ (corresponding to the post-normalization quantity in Bahadir et al.),
\begin{equation}
\tilde{m}_k \;=\; \sigma_s\!\bigl(q_k - u_k\bigr), \qquad u_k \sim \mathcal{U}(0,1),
\label{eq:loupe_relaxation}
\end{equation}
where $\sigma_s(a)=\frac{1}{1+e^{-sa}}$ is a second sigmoid (distinct from $\sigma_t$) with a (large) slope $s$ that sharpens the thresholding towards a hard step. Gradients propagate to $\phi$ through $\sigma_t$ and $\mathcal{N}_\alpha$.

Based on this formalism, Bahadir et al.\ formulated the joint sampling-reconstruction problem as an unconstrained optimization (their Eq.~6); no explicit constraint was used because the normalization layer $\mathcal{N}_\alpha$ enforces the sampling budget inside every forward pass. The resulting expectation over the stochastic mask realizations was approximated using Monte Carlo sampling.

\vspace{1em}

\bmsubsubsection*{Gumbel-softmax based sampling optimization}

An alternative, common way to implement the sampling optimization process is by (i) replacing the normalization layer $\mathcal{N}_\alpha$ used by LOUPE with a simpler multiplicative rescaling, as described in Eq.~\eqref{eq:budget_rescale} below, and (ii) replacing the uniform-noise sigmoid thresholding of Eq.~\eqref{eq:loupe_relaxation} with a Gumbel-Softmax relaxation \cite{Jang2016,ravula2023optimizing}. We adopt this implementation.

To enforce the sampling budget $B$, the per-location probabilities $p_k = \sigma(\phi_k)$ are rescaled by a single global factor,
\begin{equation}
q_k \;=\; p_k \cdot \frac{B}{\sum_{j=1}^{D} p_j + \varepsilon},
\label{eq:budget_rescale}
\end{equation}
where $\varepsilon$ is a small positive constant for numerical stability. Therefore, $\sum_{k=1}^{D} q_k = B$ by construction (up to the numerical term $\varepsilon$). This equality is imposed on the rescaled sampling weights $\mathbf{q}$, whereas the realized hard masks remain stochastic; when $\mathbf{q}$ is interpreted as a vector of valid Bernoulli probabilities, this corresponds to an expected sampling budget rather than an exact number of samples in every realized mask.

A direct binary draw based on these rescaled weights would be non-differentiable and would block the gradient with respect to $\phi$. To overcome this, the Gumbel-Softmax relaxation replaces this draw with a continuous, differentiable surrogate $z_k$:
\begin{equation}
z_k = \frac{\exp\!\bigl((\log q_k + \eta_k)/\tau\bigr)}{\exp\!\bigl((\log q_k + \eta_k)/\tau\bigr) + \exp\!\bigl((\log(1-q_k) + \eta'_k)/\tau\bigr)},
\label{eq:gumbel_softmax}
\end{equation}
where $\eta_k, \eta'_k \sim \text{Gumbel}(0,1)$ are independent noise samples and $\tau > 0$ is a temperature parameter that controls the approximation sharpness. To ensure that $\log q_k$ and $\log(1-q_k)$ in Eq.~\eqref{eq:gumbel_softmax} remain well-defined, each entry of the two-class probability vector $[\,1-q_k,\,q_k\,]$ is replaced by its maximum with a small positive constant $\varepsilon$ before the logarithm is taken. This prevents $\log(1-q_k)$ from being evaluated on a non-positive argument when the rescaling of Eq.~\eqref{eq:budget_rescale} produces $q_k \ge 1$ for over-budgeted locations. Because $z_k$ is a differentiable function of the logits $\phi_k$, gradients can flow from the loss back to the sampling parameters.

To bridge the gap between this continuous relaxation and the MRI's requirement of binary (per-readout line) acquisition, we use the straight-through (ST) estimator \cite{Jang2016}. During the forward pass, $z_k$ is discretized to a hard binary value $m_k \in \{0,1\}$; during the backward pass, the gradient of this hard sample is replaced by the gradient of the continuous relaxation $z_k$. This yields physically realizable binary masks in the forward computation while providing a surrogate gradient for optimizing the discrete sampling probabilities.

The sampling parameters $\phi$ and the reconstruction-network parameters $\theta$ are jointly optimized by solving the constrained problem
\begin{equation}
\begin{split}
\{\theta^{*}, \phi^{*}\} \;=\; \arg\min_{\theta, \phi} \ & \mathbb{E}_{\mathbf{x} \sim \mathcal{P}_{\mathbf{x}},\, M_{\phi}} \left[ \left\| f_{\theta}(A^{H} \mathbf{y}_{\phi}) - \mathbf{x}_{target} \right\|_{2}^{2} \right] \\
\text{s.t.} \ & \sum_{k=1}^{D} q_k = B
\end{split}
\label{eq:joint_single_nex}
\end{equation}
\purple{where $f_{\theta}:\mathbb{C}^{ D}\to\mathbb{R}^{D}$  is a reconstruction network which maps the complex-valued input into a real-valued magnitude image,} $\mathcal{P}_{\mathbf{x}}$ denotes the training-data distribution, $\mathbf{q}$ is the vector of rescaled sampling probabilities (defined in Eq.~\eqref{eq:budget_rescale}), and the constraint $\sum_{k=1}^{D} q_k = B$ is imposed on these probabilities. 
The expectation over $M_{\phi}$ accounts for the stochastic mask realizations. Additionally, $\mathbf{x}_{target}$  denotes the magnitude image reconstructed from the fully sampled \textit{$k$-space} data. Therefore, the $\ell_2$ loss is computed between the magnitudes of the ground-truth and reconstructed images, as in the LOUPE implementation \cite{bahadir2020deep}.

\bmsubsection{\textbf{Proposed NexOP framework}} 

\vspace{1em}

\bmsubsubsection*{Overview}

We develop a framework for jointly optimizing the sampling scheme across a multi-NEX acquisition, together with a unique architecture that reconstructs a single high-SNR image from multiple low-SNR measurements. Our framework, shown in Fig.~\ref{fig1}, consists of two main components: (i) a Sampling Module that optimizes the multi-NEX sampling operator $M_{\psi}$, and (ii) a Reconstruction Module $g_{\theta}$ that maps the low-SNR, undersampled multi-repetition \textit{$k$-space} measurements $\mathbf{y}_{\psi}$  to a single, reconstructed, high-SNR image. This module is an unrolled deep network comprising cascaded Multi-Repetition (MR) steps followed by a fusion block, and leverages shared information across the NEX dimension to improve the final image quality. The MR blocks perform both image-domain processing, which enables aliasing suppression, denoising, and SNR enhancement, and NEX-specific data consistency operations that ensure fidelity to the acquired \textit{$k$-space} data. Details are provided in the following sections.

\vspace{1em}

\bmsubsubsection*{NEX-aware sampling}  

Building on the formulation introduced in Eq.  \eqref{eq:learned_no_nex_operator} for the single-acquisition case, we define a \textit{multi-repetition} measurement process, with $NEX$ repetitions, by 
\begin{equation}
\mathbf{y}_{\psi} = M_{\psi} A\mathbf{x} + \boldsymbol{\epsilon}  
\label{eq3_new_operator} 
\end{equation} where $A$ is an extended forward operator corresponding to a multi-NEX acquisition, $M_{\psi} \in \{0,1\}^{NEX \times D}$ is a set of learnable sampling masks parameterized by $\psi$, and $NEX \times D$ is the total number of pixels in the fully sampled multi-NEX \textit{$k$-space}. The measurements $\mathbf{y}_\psi$ are thus defined for this extended domain.  

Note that in contrast to the common approach of repeating a single sampling mask across all acquisitions, our formulation allows more freedom: here, the samples can be freely distributed across both the spatial \textit{$k$-space} dimensions and the $NEX$ domain. Moreover, our framework does not impose fixed per-repetition sampling rates and allows the \textit{acceleration factor} to vary across acquisitions; to our knowledge, this specific NEX-adaptive sampling-optimization formulation has not been previously explored.

\bmsubsubsection*{\textbf{Joint multi-NEX sampling-reconstruction optimization}}


We jointly optimize the parameters of the Sampling Module $M_{\psi}$ and those of the Reconstruction Module $g_\theta$  by solving:

\begin{equation}
\begin{split}
\{\theta^{*}, \psi^{*}\} = \arg\min_{\theta, \psi} \ & \mathbb{E}_{\mathbf{x} \sim \mathcal{P}_{\mathbf{x}},\, M_{\psi}} \left[\left\|  g_{\theta}(A^{H}\mathbf{y}_{\psi}) - \mathbf{x}_{target} \right\|_{2}^{2}\right] \\
\text{s.t.} \ & \sum_{k \in \Omega_{\mathrm{optim}}} q_k = B
\end{split}
\label{eq:joint_obj}
\end{equation} 

where $g_{\theta}:\mathbb{C}^{NEX \times D}\to\mathbb{R}^{D}$  is the Reconstruction Module 
(Fig. \ref{fig1}), \purple{ which maps the complex-valued multi-repetition input into a single real-valued magnitude image.} $\mathcal{P}_{\mathbf{x}}$ now represents the distribution of a multi-repetition training dataset, $\mathbf{q}$ is the vector of rescaled sampling weights used to construct the stochastic masks\purple{, and $\Omega_{\mathrm{optim}}$ denotes the set of optimizable $k$-space locations constrained by the budget $B$, where $|\Omega_{\mathrm{optim}}| = N_{\text{optim}}$}. The expectation over $M_{\psi}$ accounts for the stochastic masks drawn through the Gumbel-Softmax relaxation during training. The adjoint operator $A^H$ operates on data acquired from each repetition individually, transforming the multi-repetition \textit{$k$-space} measurements into a stack of images. The constraint $\sum_{k \in \Omega_{\mathrm{optim}}} q_k = B$ enforces the budget on the rescaled sampling weights, whereas the realized binary masks remain stochastic. In the Bernoulli interpretation, this corresponds to an expected sampling budget when the entries of $\mathbf{q}$ are valid probabilities.

The target image is defined as the mean \purple{of the NEX-specific magnitude images computed from the fully sampled k-space data:}     $\mathbf{x}_{target}=\frac{1}{NEX} \sum_{n=1}^{NEX}|\mathbf{x}_n|$. This encourages the network to learn the mapping from multiple low-SNR aliased images into a final, high-SNR alias-free image.  


\begin{algorithm*}
\caption{NexOP forward pass}\label{algo:forward}
\begin{algorithmic}[1]
\Require
\Statex Fully sampled $NEX$ repetitions  \(\mathbf{x}_n, \quad n=1,...,NEX\)
\Statex Sampling budget \(B\)
\Statex Auto-calibration signal indices \(\mathrm{ACS\_idx}\)
\Ensure Reconstructed image \(\hat{\mathbf{x}}\), binary masks \(M_\psi\)
\State \(\mathbf{p} \leftarrow \sigma(\psi)\) \Comment{Logits to initial probabilities via Sigmoid}
\State \(\mathbf{q} \leftarrow \mathbf{p} \cdot \frac{B}{\sum p + \varepsilon}\) \Comment{Normalize probabilities}
\State \(z \leftarrow \mathrm{GumbelSoftmax}(log(\mathbf{q}))\) \Comment{Draw [0,1] sample probabilities}
\State Reshape \(z\) into \(NEX\) distinct repetition masks \(m_1, \dots, m_{NEX}\)
\State \(m_1[\mathrm{ACS\_idx}] \leftarrow 1\) \Comment{Ensure ACS sampling}
\State \(M_\psi \leftarrow \mathrm{Stack}(m_1, \dots, m_{NEX})\) 
\State \(\mathbf{y}_\psi \leftarrow M_\psi A[\mathbf{x}_1, ..., \mathbf{x}_{NEX}]\) \Comment{Undersample all NEX}
\State \(\hat{\mathbf{x}} \leftarrow g_{\theta}(A^H \mathbf{y}_\psi)\) \Comment{Pass through the reconstruction network}
\label{algorithm1}
\end{algorithmic}
\end{algorithm*}

\bmsubsection{NexOP  implementation}

\vspace{1em}

\bmsubsubsection*{Proposed multi-NEX sampling optimization method}\label{sec_NexOP}

\textbf{Sampling budget definition}.
For both NexOP and the competing methods (described in Section \ref{benchmarking}), we assume that an auto-calibration signal (ACS) area, which is a fully sampled area in the center of \textit{$k$-space} consisting of $N_{\text{ACS}}$ locations, is sampled in the first repetition (i.e. that of $NEX=1$). This area is commonly acquired for sensitivity maps estimation, e.g., using the ESPIRiT algorithm \cite{Uecker2014}, which is required for processing multi-coil data. Because this first-repetition ACS acquisition is fixed, those pixels are excluded from the first-mask optimization; ACS locations in later repetitions may be selected by the learned sampling process. Therefore, the number of candidate \textit{$k$-space} locations that can be selected during the optimization from the set $\Omega_{\mathrm{optim}}$ is given by $N_{\text{optim}}=(NEX \times D)-N_{\text{ACS}}$. 

Additionally, we define $B$ as an optimizable sampling budget, i.e, the goal of our framework is to find an optimal strategy for distributing the $B$ samples across the $N_{\text{optim}}$ available locations.  

The acceleration factor $R$, defined as the overall acceleration across all repetitions, accounts for both the number of sampled locations in the ACS region, $N_{\text{ACS}}$, and the number of samples allocated by our framework, $B$. Accordingly, these quantities are related by:
\begin{equation}
R = \frac{NEX \times D}{B+N_{\text{ACS}}}.
\end{equation}

\textbf{Sampling optimization implementation.} The framework optimizes the acquisition pattern by learning a set of unconstrained real-valued parameters (logits), denoted as $\psi \in \mathbb{R}^{N_{optim}}$. These logits correspond to the $N_{\text{optim}}$ candidate \textit{$k$-space} locations, excluding the ACS region. The logits are first mapped to the $(0,1)$ range via a sigmoid activation function, $\sigma(\cdot)$, to produce initial probability values $p$.  This constraint is enforced similarly to the LOUPE implementation (Section \ref{sec_LOUPE}): these probabilities are rescaled to match the non-ACS sampling budget $B$ via the operation $q = p \cdot \frac{B}{\sum p + \epsilon}$, where $\epsilon$ is a small constant ensuring numerical stability. This normalization yields the final constrained sampling probabilities $q$, which serve as the parameters for the subsequent differentiable sampling process.

To accurately simulate a physically realizable MRI acquisition during the forward pass, these continuous probabilities $q$ are converted into discrete binary masks. To maintain end-to-end differentiability, we utilize the Gumbel-Softmax relaxation \cite{Jang2016} to draw samples based on the learned probabilities, such that $z = \mathrm{GumbelSoftmax}(\log(q))$; the intermediate variable $z$ is then reshaped into distinct sampling masks $m_1, \dots, m_{NEX}$. The ACS region is then explicitly added to the sampling mask of the first repetition, $m_1$. The resulting set of masks constitutes the multi-NEX operator $M_\psi$, which is used to retrospectively undersample the data. As described in Eq.~\eqref{eq:joint_obj} and Algorithm 1, the sampling parameters $\psi$ are optimized in tandem with the reconstruction weights $\theta$ in a fully end-to-end fashion.

\bmsubsubsection*{Reconstruction Module architecture}

Our Reconstruction Module $g_{\theta}$ is designed to learn the mapping from multi-repetition, low-SNR, undersampled measurements into a single, high-SNR reconstructed image. Rather than receiving a single averaged volume, the network receives a concatenated input comprising the zero-filled reconstructions of all repetitions, denoted by $\mathbf{A}^H \mathbf{y}_\psi$, and their corresponding sampling masks. This allows the model to exploit the diverse \textit{$k$-space} coverage that may exist across different repetitions. Our architecture adopts an unrolled optimization strategy, which consists of interleaved image-domain neural networks with data consistency (DC) that ensure consistency to the acquired measurements, as it has demonstrated state-of-the-art performance in MRI reconstruction \cite{aggarwal2018modl, Hammernik2017LearningAV, hammernik2021systematic, johnson2023deep, wang2022high, pramanik2023adapting}.

However, conventional unrolled methods are traditionally limited to processing a single \textit{$k$-space} acquisition \cite{aggarwal2018modl,hammernik2023physics}. To overcome this, we introduce a multi-NEX unrolled architecture composed of cascaded Multi-Repetition data processing blocks, denoted as \textit{MR Steps} (Fig. \ref{fig1}a). We describe this architecture here for the case of three repetitions; an extension to a different number of repetitions is straightforward. Each \textit{MR Step} incorporates the following internal data processing sequence to leverage shared information while maintaining repetition-specific fidelity. First, the input is split into its magnitude and phase components. The magnitude part is fed into an image-domain neural network, $NN_{\theta}$, which extracts complementary features from the information shared across all repetitions to perform aliasing removal, denoising, and SNR enhancement. Specifically, $NN_{\theta}$ is a 5-layer residual convolutional neural network composed of $3 \times 3$ convolutional layers with 64 feature channels, batch normalization, and ReLU activations, followed by a global residual connection. The real-valued output of  $NN_{\theta}$ is then multiplied back with the original phase for each repetition separately to restore the complex-valued multi-NEX data. Subsequently, these data are processed by NEX-specific data consistency (DC) blocks, which utilize a physics-based model of the forward process that operates in \textit{$k$-space}, and apply a DC step using a conjugate gradient optimizer \cite{aggarwal2018modl}. The DC constraints are applied to each repetition separately, to ensure fidelity to the acquired measurements. The outputs of these DC blocks are concatenated to form the output of the \textit{MR Step} block, which serves as the input for the subsequent block (Fig. \ref{fig1}a). In the final iteration, we omit the DC step, thereby mitigating noise reinforcement, which is a significant limitation in low-field MRI measurements.

Finally, after five cascaded blocks, the data are passed through a Fuser block (Fig. \ref{fig1}a), which concatenates the multi-repetition data and computes the mean of the \purple{NEX-specific} magnitude images, producing a single, high-SNR reconstructed image. This image, denoted by $\hat{x}=g_{\theta}(A^{H}\mathbf{y}_{\psi})$, is the output of the framework; during training it is compared to $\mathbf{x}_{target}$, as described in Eq.~\eqref{eq:joint_obj}.  

The sampling and reconstruction modules are trained jointly (Eq.~\eqref{eq:joint_obj}). Our modular design hence enables the framework to learn highly efficient acquisition strategies in tandem with a reconstruction process tailored specifically to the multi-NEX setting.

\bmsubsection{Data}
Experiments were performed with the M4Raw database \cite{Lyu2023}, which includes data acquired with a $0.3\,\mathrm{T}$ low-field scanner (Oper-0.3, Ningbo Xingaoyi), with $NEX=3$ repetitions per scan. We divided this database into two contrast-specific subsets: (i) a T1-weighted (T1w) subset, acquired with a Spin Echo pulse sequence ($TE=18.4\,\mathrm{ms}$, $TR = 500\,\mathrm{ms}$); and (ii) a T2-weighted (T2w) subset, acquired with a Fast Spin Echo sequence ($TE=128\,\mathrm{ms}$, $TR = 5500\,\mathrm{ms}$).  Each subset was divided into subject-specific groups consisting of 128/25/30 subjects for training/validation/test. 

All scans were acquired using a four-channel head coil; we computed the coil-specific sensitivity maps using the ESPIRiT algorithm \cite{Uecker2014}. 18 slices were acquired per subject, and per repetition, using 2D multi-slice protocols with an in-plane resolution of $0.94 \times 1.23\,\mathrm{mm}^2$,  $5\,\mathrm{mm}$ slice thickness, a $1\,\mathrm{mm}$ gap between adjacent slices, matrix size of $[256 \times 195]$ [H, W], and a field-of-view of $240 \times 240\,\mathrm{mm}^2$.

\bmsubsection{Benchmarking against competing methods}\label{benchmarking}

We benchmarked NexOP against five methods for selecting the sampled \textit{$k$-space} locations, including both fixed heuristic sampling schemes, such as variable density, and learning-based sampling optimization frameworks. To the best of our knowledge, existing approaches do not explicitly exploit the NEX dimension for sampling optimization. Therefore, to enable a fair comparison with  DL-based methods under multi-acquisition settings, we introduce extensions of the well-established LOUPE framework \cite{bahadir2020deep}, which operate across multiple repetitions, and benchmark NexOP against them.
\begin{enumerate}
    \item \textbf{Variable-Density (VD) Poisson-disc undersampling}: this is a standard method, applied only to a single acquisition ($NEX=1$). It was implemented using the SigPy toolbox \cite{sigpy}.
    
    \item \textbf{Multi-NEX Variable-Density (Multi-NEX VD) undersampling}: This method uses three independent, pre-defined VD Poisson-disc sampling masks for the three repetitions, as suggested by Schoormans et al.~\cite{Schoormans2020}. It therefore exploits the NEX dimension but relies on fixed, non-learnable sampling masks. Those masks were also generated using SigPy \cite{sigpy}.
    
    \item \textbf{LOUPE}: a DL-based sampling optimization method developed by Bahadir et al., \cite{bahadir2020deep}, described above in Section~\ref{sec_LOUPE}. As LOUPE was originally formulated for a single acquisition ($NEX=1$), it was applied here to the first repetition only, without extending the learned sampling pattern to additional repetitions.
    
    \item \textbf{LOUPE-ext ($NEX=2$)}: an extension of LOUPE in which a single optimized sampling mask is learned and applied across the first and second repetitions, utilizing the learning framework described in Section~\ref{sec_LOUPE}. This method does not apply sampling in the third repetition.
    
    \item \textbf{LOUPE-ext ($NEX=3$)}: an extension of LOUPE in which a single sampling mask is learned and applied across all three repetitions. This method also follows the framework described in Section~\ref{sec_LOUPE}.
\end{enumerate}


For a fair comparison, all methods were assigned the same task: optimizing a total budget of $B$ samples across three repetitive acquisitions. All methods were required to acquire an ACS region of $20 \times 20$ pixels at the center of \textit{$k$-space} in the first repetition; as some methods repeat identical sampling across NEX acquisitions, in that case the budget $B$ was adjusted to account for the repetitive ACS acquisitions and satisfies $R = \frac{NEX \times D}{B+ NEX \cdot N_{\text{ACS}}}$. Additionally, to isolate the effect of sampling design and ensure comparable reconstruction performance, all methods were implemented using the proposed multi-NEX Reconstruction Module.

\bmsubsection{Computational implementation.}

The proposed NexOP framework and all competing methods were implemented in Python using the PyTorch deep-learning library \cite{paszke2019pytorch}. Our implementation of all LOUPE-based methods followed the process described in Section \ref{sec_LOUPE}. 
 
To ensure a fair comparison, all the reconstruction networks were trained following a consistent protocol. Specifically, the network weights were optimized using an $\ell_2$ loss function and the Adam optimizer \cite{kingma2014adam}. The optimization process utilized a step learning rate scheduler that halved the learning rate every 14 epochs. The initial learning rates for the reconstruction networks were empirically calibrated for each experimental setting (e.g., data subsets and acceleration factors); typical values were in the range of $10^{-4}$ to $10^{-6}$. For all methods involving differentiable sampling optimization (NexOP, LOUPE, and the LOUPE extensions), the non-differentiable Bernoulli mask draw of Eq.~\eqref{eq:joint_single_nex} was relaxed via the Gumbel-Softmax estimator (Eq.~\eqref{eq:gumbel_softmax}) combined with the straight-through (ST) variant \cite{Jang2016}: the forward pass uses hard binary samples, while gradients in the backward pass are propagated through the continuous relaxation. The temperature $\tau$ was annealed during training from $\tau=1.0$ by a factor of $0.95$ per epoch (with a floor of $\tau=0.1$), and fixed to $\tau=0.5$ at test time. The slope parameter $s$ of LOUPE's sigmoid-thresholding relaxation in Eq.~\eqref{eq:loupe_relaxation} was not used, since the Gumbel-Softmax relaxation was adopted uniformly across all methods.

For the VD and Multi-NEX VD methods, which utilize fixed sampling patterns, this training protocol was applied to optimize the unrolled reconstruction network \purple{and evaluate the sampling-strategy effects.}

In contrast, the proposed NexOP, LOUPE, and LOUPE extensions frameworks undergo a joint training process where both the sampling mask layer and the unrolled Reconstruction Module were optimized end-to-end. To ensure stable convergence between these two fundamentally different components, we used different learning rates by applying a higher learning rate to the sampling mask parameters, typically ranging from $10^{-2}$ to $10^{-4}$, and calibrated for each experimental setting. All models were trained for a total of 30 epochs using a batch size of 16. Moreover, to maintain stability and prevent gradient divergence during optimization, gradient clipping was applied independently to the parameters of both the reconstruction and sampling operators. The computations were performed on NVIDIA L40S GPU.

\textbf{Data processing}. All data processing and transformations were implemented using the \textit{fastMRI} open-source package \cite{zbontar2018fastmri}. Multi-coil \textit{$k$-space} data processing utilized the ESPIRiT-based sensitivity maps estimation \cite{Uecker2014}, as implemented in SigPy \cite{sigpy}. To enable processing with real-valued networks, complex-valued data were represented by concatenating the real and imaginary components along the channel dimension. 

\begin{figure*}[b]
\centering
\includegraphics[width=1\textwidth]{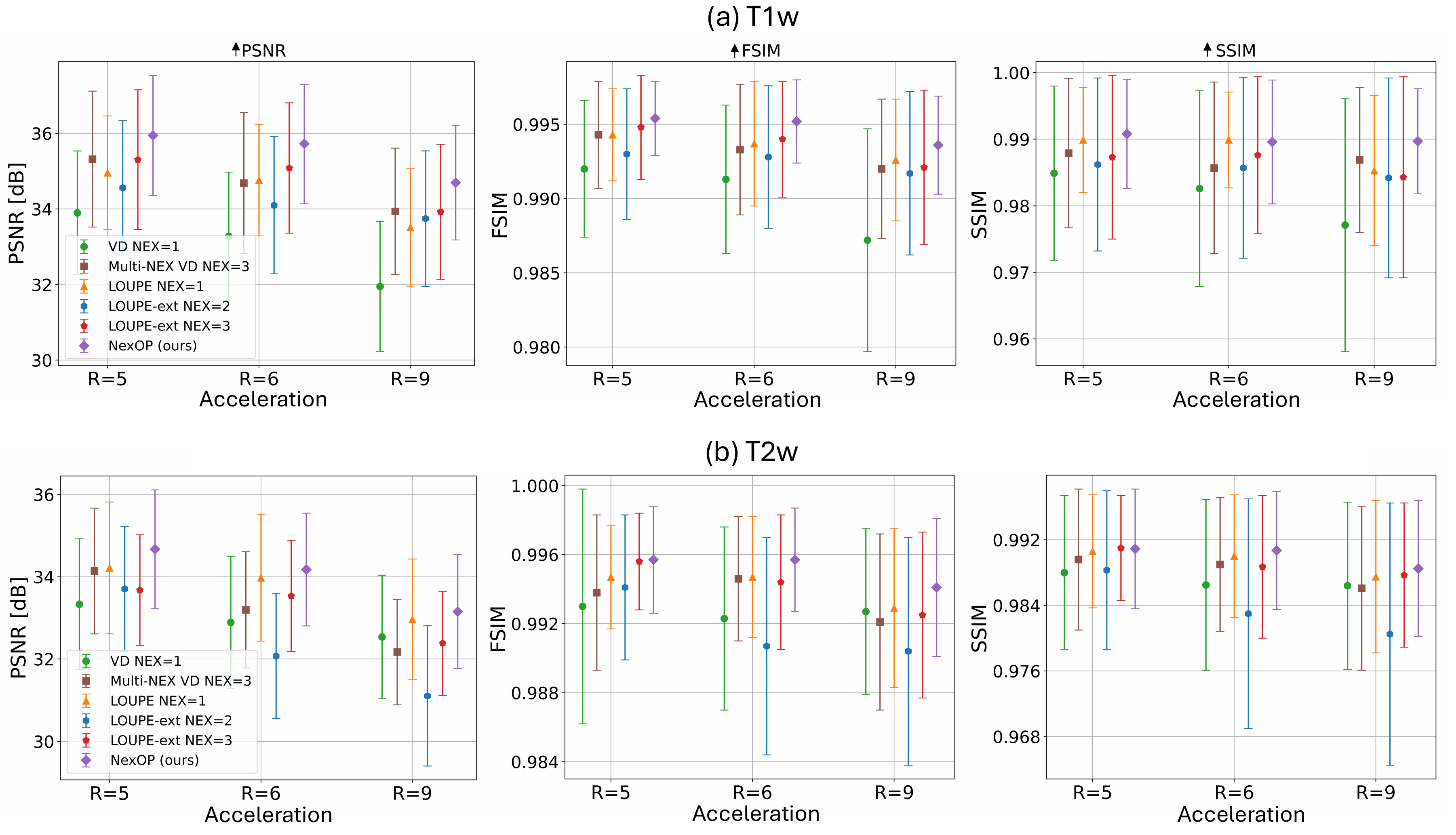}
\caption{\textbf{Quantitative comparison of sampling optimization strategies for different acceleration factors ($R=5, 6, 9$)}. The proposed NexOP framework (purple diamond) is compared to five other methods: VD, Multi-Nex VD, LOUPE, and the two LOUPE extensions. Results are presented for two subsets: (a) T1w and (b) T2w data. NexOP consistently achieves superior PSNR, SSIM, and FSIM scores. This performance comparison highlights the benefit of joint NEX and sampling optimization.}\label{fig2_metrics}
\end{figure*}

\bmsubsection{Evaluation methodology}
To comprehensively assess the performance of the proposed joint sampling and reconstruction framework, we employed both qualitative and quantitative evaluation strategies. Qualitatively, we performed a visual assessment of the reconstructed images to evaluate the preservation of fine anatomical details, tissue contrast, and the suppression of noise and aliasing artifacts compared to the ground truth. 

Quantitatively, the reconstruction fidelity was measured against \purple{the previously defined $\mathbf{x}_{target}$} using three standard full-reference image quality metrics: Peak Signal-to-Noise Ratio (PSNR), Structural Similarity Index Measure (SSIM) \cite{Wang2004}, and Feature Similarity Index Measure (FSIM) \cite{Zhang2011}, which emphasizes the preservation of low-level structural features that are highly critical in medical imaging diagnostics \cite{mason2019comparison}. All quantitative metrics were computed exclusively within a region of interest (ROI) containing the relevant anatomy, masking out the background. These ROI masks were created automatically using an intensity thresholding approach followed by morphological operations, specifically binary closing and hole filling, utilizing the SciPy library \cite{virtanen2020scipy}. This morphological processing ensured that all image quality metrics accurately reflected the reconstruction fidelity of the ROI only.



\section{Results}\label{sec3}

To evaluate the proposed NexOP framework, we conducted a comprehensive series of experiments comparing its performance against established \textit{$k$-space} sampling optimization and retrospective undersampling techniques. In the following sections, we first present a quantitative analysis across varying acceleration factors to validate the stability of our approach. Subsequently, we provide detailed qualitative assessments of the reconstructed T1w and T2w image sets. Finally, we visualize and examine the underlying probability maps and sampling masks learned by our joint optimization pipeline.

\bmsubsection{\textbf{Quantitative results}}

We compared the proposed NexOP framework against the five baseline methods in experiments with an identical sampling budgets, as detailed in Section \ref{benchmarking}. These evaluations were performed across different accelerations of $R=5,6,9$. We quantitatively evaluated the results using the PSNR, SSIM, and FSIM metrics.

The results (Fig. \ref{fig2_metrics}) demonstrate that NexOP consistently achieves the highest image quality metrics across all acceleration factors, for both the T1w and T2w data. This consistency demonstrates that optimizing both the sampling pattern and the reconstruction process across the NEX dimension leads to superior image quality \purple{compared to the reference methods}. Furthermore, the relatively narrow STD bars that characterize the NexOP results (Fig. \ref{fig2_metrics}) indicate low variability across the evaluated test subjects.

\begin{figure*}[]
\centering
\includegraphics[width=1\textwidth]{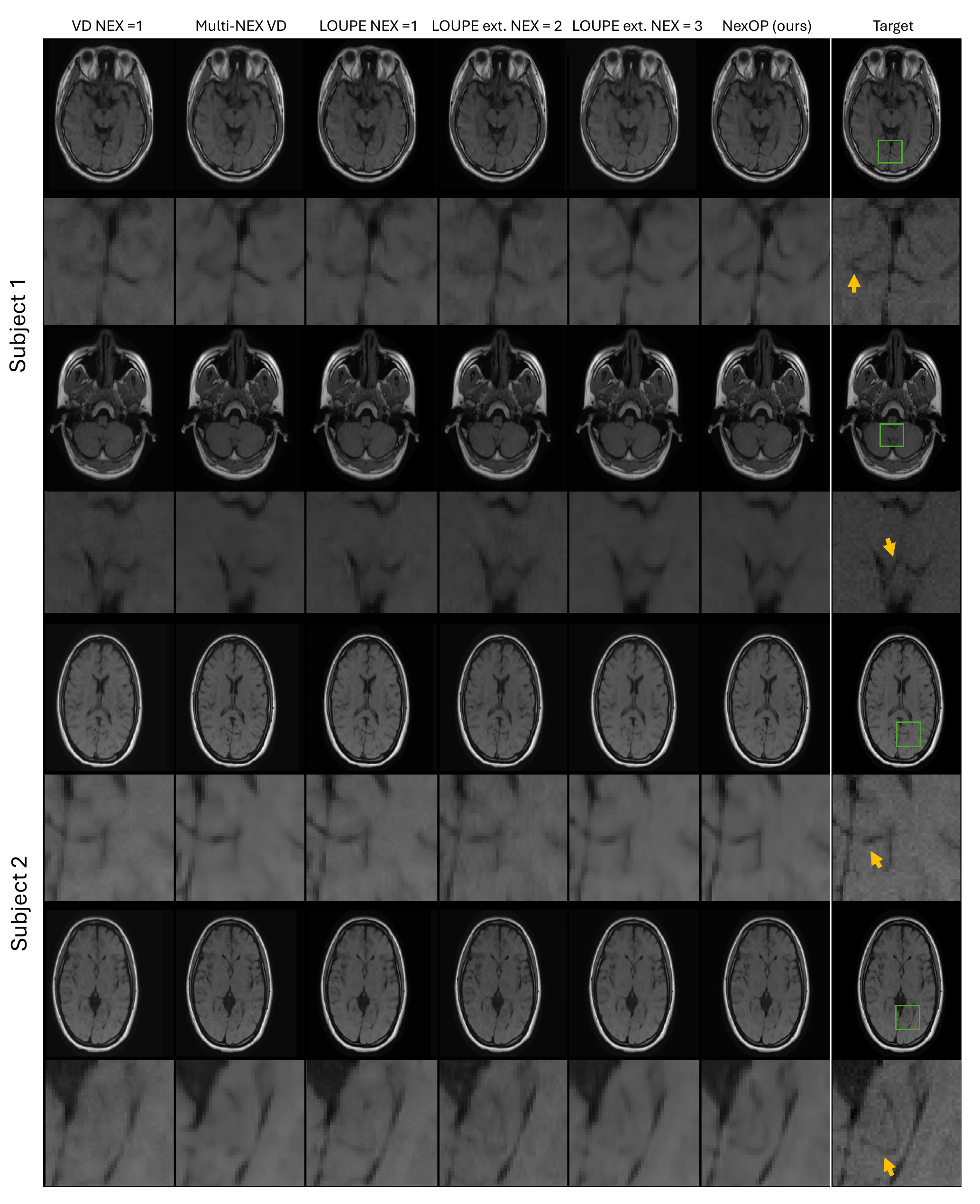}
\caption{\textbf{Visual comparison of T1w reconstructions for an acceleration factor of $R=6$.} This figure presents results from two test subjects (Subject 1, top; Subject 2, bottom). We compare our proposed NexOP method against the Variable-Density (VD $NEX=1$, Multi-NEX VD) and LOUPE ($NEX=1$, ext. $NEX=2$, ext. $NEX=3$) methods. Orange arrows in the target images indicate anatomical features for evaluation.}\label{fig5_T1_visual}
\end{figure*}
\begin{figure*}[b]
\centering
\includegraphics[width=1\textwidth]{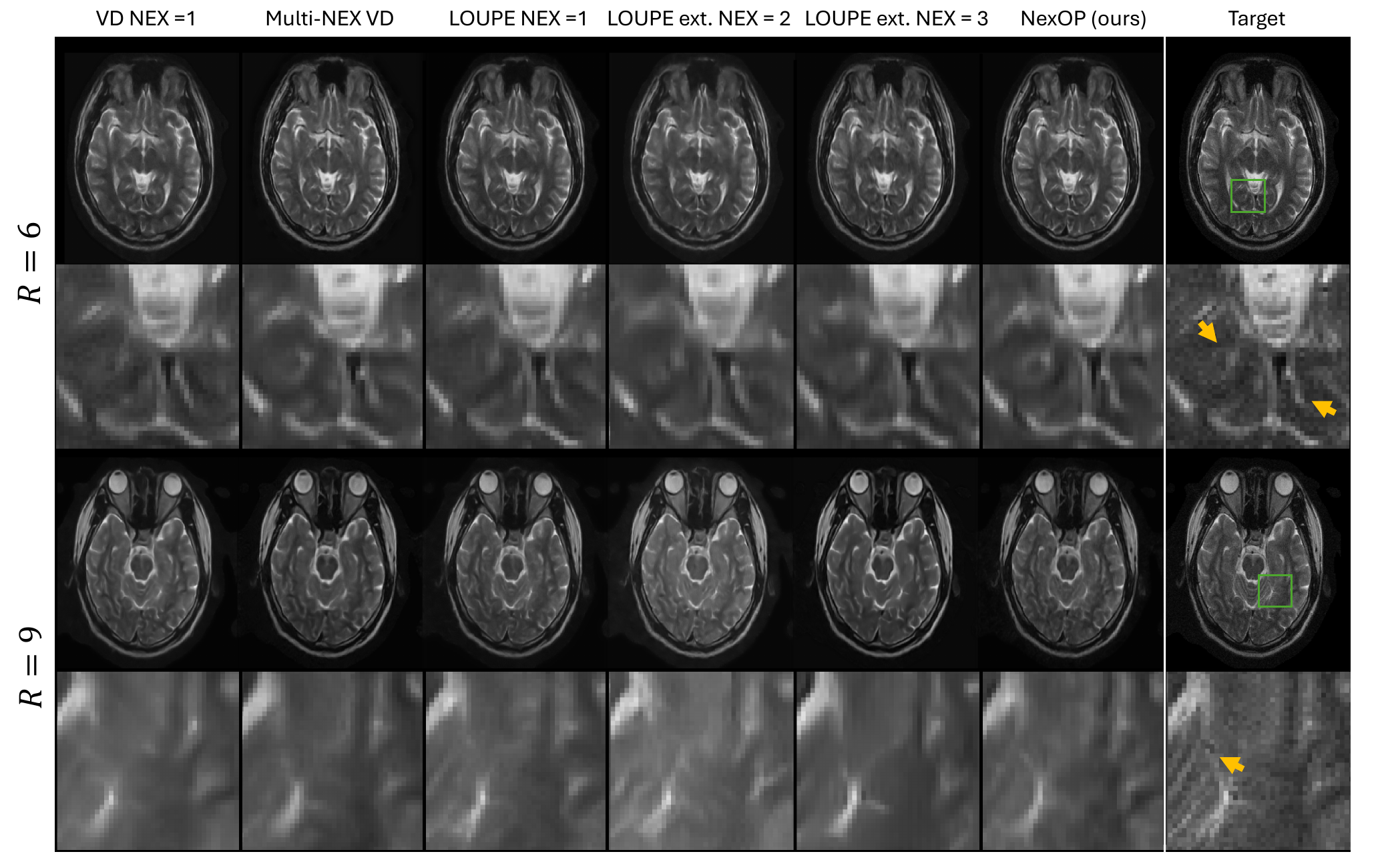}
\caption{\textbf{Qualitative comparison of T2w reconstructions for acceleration factors $R=6$ and $R=9$.} This figure compares our method against the five baseline methods (VD, Multi-NEX VD, LOUPE, and LOUPE extensions) at $R=6$ (top rows) and $R=9$ (bottom rows). The results indicate that NexOP effectively preserves fine details (orange arrows in the target image), suppresses noise, and demonstrates robust performance even for the challenging $R=9$ acceleration.}\label{fig6_T2_visual}
\end{figure*}

\bmsubsection{\textbf{Qualitative assessment}}

\purple{Next, we present side-by-side images reconstructed by the different methods for T1w and T2w data.}

\textbf{T1-weighted data}. Figure \ref{fig5_T1_visual} presents representative reconstructions for the T1w subset for an acceleration factor of $R=6$. 
The results indicate that the VD-based and LOUPE-based methods produce images exhibiting over-smoothing and blurring, with a loss of fine anatomical textures. Additionally, some of these methods (e.g., the VD and LOUPE-ext $NEX=2$ methods) yield noisy reconstructions. In contrast, the proposed NexOP framework produces higher-quality images, with improved structural fidelity to the target image. By learning distinct, complementary sampling patterns along the repetition dimension, NexOP successfully reconstructs fine anatomical details (as indicated by the arrows) while suppressing background noise.

\textbf{T2-weighted data}. To further demonstrate the proposed framework, Fig.~\ref{fig6_T2_visual} presents reconstruction results for experiments with the T2w subset at acceleration factors of $R=6$ and $R=9$. Consistent with the T1w results, the baseline methods exhibit lower image quality compared to NexOP. For example, at $R=6$, the single-NEX techniques (VD $NEX=1$ and LOUPE $NEX=1$) yield overly smooth images with substantial loss of fine details. The Multi-NEX VD and the two LOUPE-ext methods recover slightly more image \purple{details but miss some small structural details, as indicated by the orange arrows.}

\bmsubsection{\textbf{Demonstration of learned probability maps and sampling masks}}
 
\begin{figure}[htbp]
    \centering
    \includegraphics[width=0.7\textwidth]{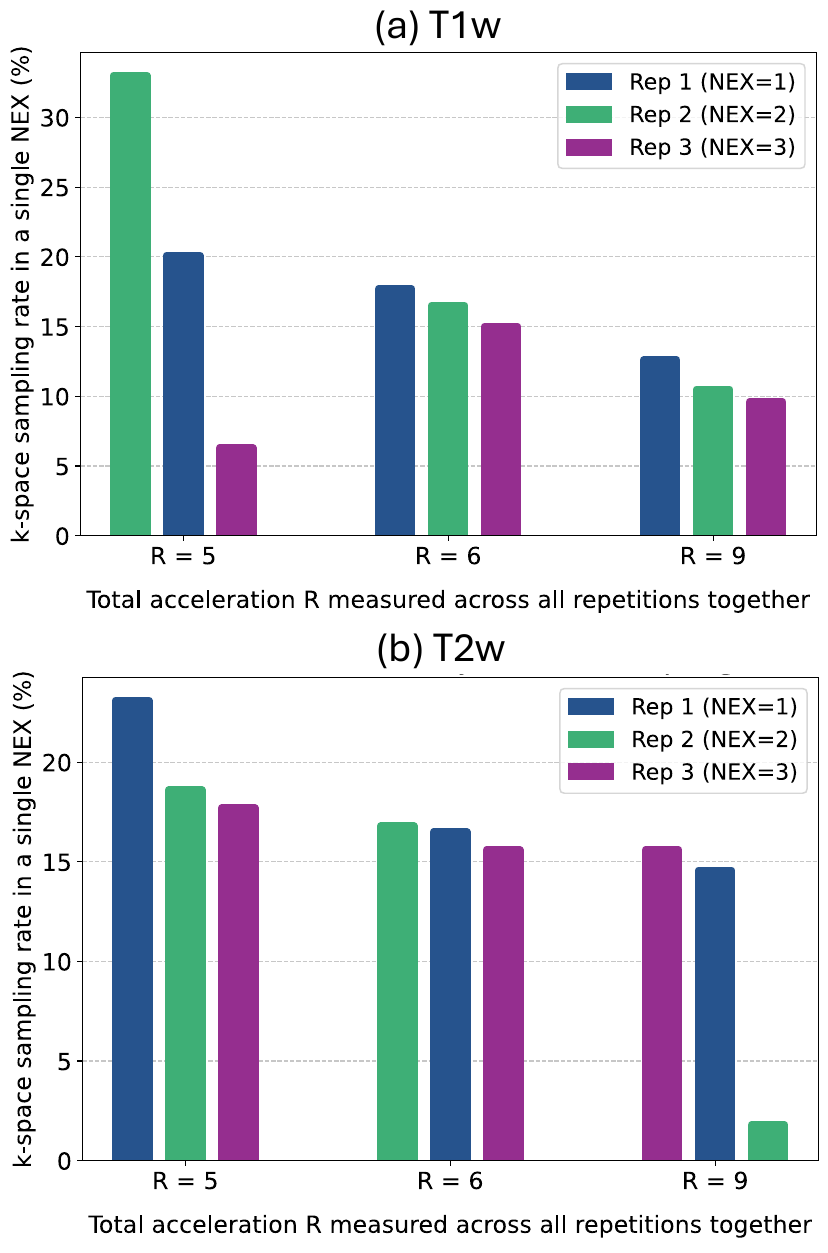}
    \caption{\textbf{Distribution of \textit{$k$-space} sampling rates across repetitions.} The percentage of sampled \textit{$k$-space} locations for each of the three repetitions ($NEX=1,2,3$) for (a) T1w and (b) T2w data. The sampling rate on the vertical axis is defined relative to a single fully sampled acquisition, whereas the horizontal axis indicates the total acceleration factor $R$ across the multi-NEX domain. \purple{Notably, NexOP yields non-uniform, \textit{progressively decreasing}} sampling strategies across NEX, where a larger fraction of samples is allocated to earlier acquisitions and a smaller fraction to subsequent repetitions. \purple{These results suggest that NexOP leverages the flexibility in distributing sampling across the repetitions domain to optimize the final image quality.}}
    \label{fig5:Per_NEX_Sampling}
\end{figure}

To analyze the acquisition strategies optimized by NexOP, we examined the distribution of the learned \textit{$k$-space} sampling rates across the repetition dimension. Figure \ref{fig5:Per_NEX_Sampling} presents the percentage of \textit{$k$-space} samples within each of the three repetitions relative to a single fully sampled acquisition, vs. $R$, which is the acceleration across all repetitions together. This analysis is performed for both T1w and T2w data, for different acceleration factors, without including the ACS. 

The results (Fig. \ref{fig5:Per_NEX_Sampling}) indicate that the NexOP framework adaptively allocates the budget in a non-uniform and contrast-dependent manner. For example, in the T1w and $R=5$ case, the optimized strategy allocates about 30\% of the samples to the first repetition, approximately 20\% to the second, and only about 6\% of \textit{$k$-space} is sampled in the last repetition. Another strategy was learned for T2w data at $R=9$, in which the framework assigns nearly equal sampling rates of approximately 15\% to the first and second repetitions, and less than  5\% to the last repetition. Yet, the sampling rate is monotonically decreasing along the NEX dimension;  these results suggest that the network leverages information from earlier acquisitions to inform and optimize later repetitions.

\begin{figure*}[]
\centering
\includegraphics[width=1\textwidth]{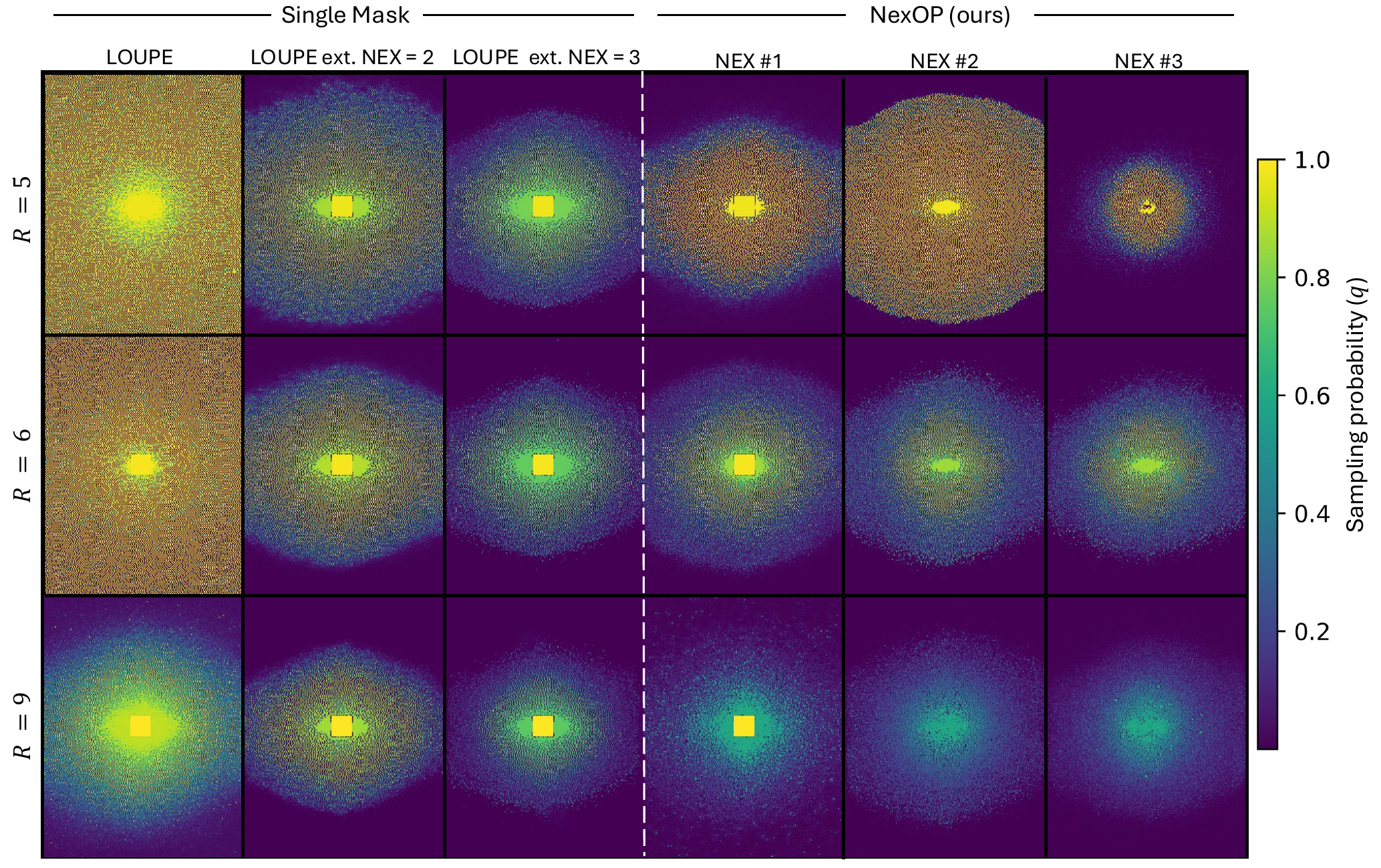}
\caption{\textbf{Visualization of the learned numerical sampling probabilities.} This figure illustrates the optimized probability maps for the T1w subset, denoted as $q$, with the colorbar indicating the numerical sampling probability. Notably, the NexOP framework learns probabilities that vary across repetitions: in most settings, the first repetition covers a broad range of spatial frequencies, whereas later repetitions tend to concentrate their sampling probability near the \textit{$k$-space} center.}\label{fig6_maps}
\end{figure*}

Figure \ref{fig6_maps} complements the statistical analysis by presenting the learned sampling probability maps $q$ (which were defined in Sections \ref{sec_LOUPE} and \ref{sec_NexOP}) across varying acceleration factors ($R=5, 6, 9$) for the T1w subset. In these visualizations, the color scale represents the probability of sampling each individual \textit{$k$-space} location. While the budget-constrained normalization can numerically produce values exceeding $1$, these are effectively treated as a probability of $1$ during the sampling process and are clipped to the $[0, 1]$ range for visualization. Because each LOUPE-based method learns only a single sampling mask, their learned probability maps are displayed in a single column per method. In contrast, the proposed NexOP learns distinct sampling maps and hence different probabilities across repetitions; those are presented in three columns. \purple{In particular, the NexOP framework learns probabilities that vary between repetitions. Typically, in most settings, the first repetition covers a wide range of spatial frequencies, whereas later repetitions tend to concentrate their sampling near the center of \textit{$k$-space}.}

\purple{Another aspect of the learned sampling strategy is demonstrated in Figure \ref{fig7_masks}, which presents the accumulated sampling masks.} The maps represent the total number of times each \textit{$k$-space} location is acquired across the entire scan. The baseline methods, such as LOUPE ext. $NEX=3$ optimize an accumulation by repeating an identical mask across all acquisitions. Conversely, the NexOP framework (Fig. \ref{fig7_masks}, right column) yields an optimized variable-density distribution. This effectively extends the VD property into the NEX dimension, where lower spatial frequencies are sampled with higher repetition rates, whereas higher frequencies are acquired more sparsely across the scan with fewer repetitions. A further analysis supporting this observation is presented in Appendix~\ref{secA_maps_analysis}.

\section{Discussion}\label{sec_discussion}

Low-field MRI operates in an SNR-straved regime, hence multiple repetitions (NEX) are often acquired to recover adequate SNR; consequently, scan duration is often prolonged, limiting the clinical utility of this emerging modality. Although a large body of work has addressed undersampling and reconstruction for accelerated MRI \cite{bahadir2020deep,Aggarwal2020,Weiss2021,chaithya2022optimizing,ravula2023optimizing,Gautam2025}, these efforts have primarily targeted single-acquisition settings at high field, and the NEX dimension has remained largely unexploited. Here we introduced NexOP, a framework that jointly optimizes a multi-NEX \textit{$k$-space} sampling strategy and a DL-based reconstruction architecture tailored to multi-NEX data. Under a fixed acquisition budget, NexOP optimizes the allocation of \textit{$k$-space} samples across repetitions and reconstructs a single high-SNR image from multiple low-SNR measurements.

\bmsubsection{Main observations}

We benchmarked several frameworks for distributing samples across the NEX dimension, while keeping the reconstruction architecture identical (Figs.~\ref{fig2_metrics},~\ref{fig5_T1_visual},~\ref{fig6_T2_visual}); consequently, the performance differences observed in our experiments primarily reflect how each method distributes the acquisition budget across repetitions. This comparison yielded several observations. First, simply replicating an optimized mask across repetitions, as in the extended LOUPE methods, improves SNR but yields only a marginal gain over single-acquisition LOUPE (Figs. \ref{fig2_metrics}, \ref{fig5_T1_visual}, and \ref{fig6_T2_visual}), even when the sampling is optimized via deep learning. \purple{Second, simply diversifying sampled \textit{$k$-space} locations across repetitions is not inherently superior. Multi-NEX VD, which distributes samples non-uniformly, competes with, and occasionally underperforms LOUPE extensions, despite their use of fixed masks across repetitions. Both approaches still fall short of NexOP because they rely on pre-defined sampling schemes rather than learning them directly from the data.} The proposed NexOP framework, which both distributes samples non-uniformly across repetitions and optimizes the distribution in a data-driven manner, achieves the highest reconstruction quality across all experiments. Moreover, its consistent advantage across diverse tissue contrasts and acceleration factors suggests that the framework adapts its sampling strategy to the underlying tissue contrast and the acquisition setting. Taken together, these results support a central conclusion: in the low-SNR regime, the distribution of samples across the NEX dimension is a degree of freedom that, when optimized jointly with reconstruction, yields consistent gains compared to optimizing the spatial \textit{$k$-space} pattern alone.

\begin{figure*}[]
\centering
\includegraphics[width=1\textwidth]{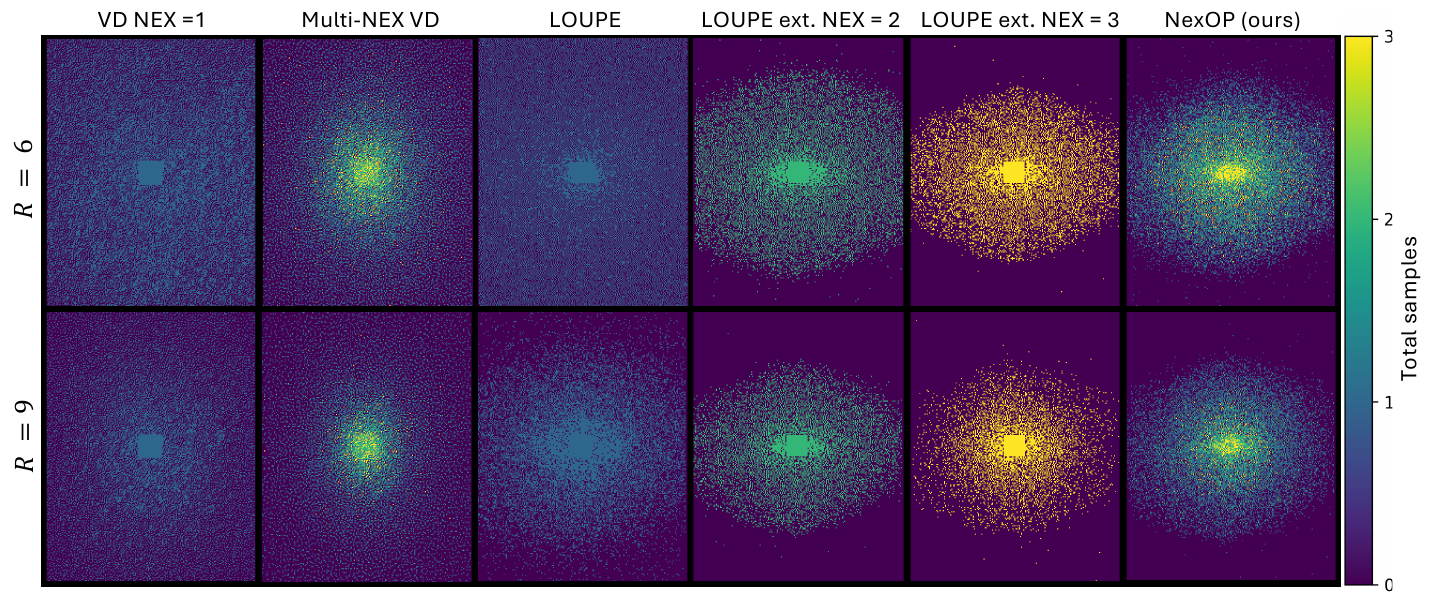}
\caption{\textbf{Visualization of accumulated sampling masks across the NEX dimension for T1w data and different acceleration factors ($R=6, 9$).} 
The accumulation maps illustrate the total number of samples at each \textit{$k$-space} location, calculated by aggregating the binary masks across all repetitions. The baseline methods VD, \textit{LOUPE}, and its extensions (e.g., \textit{LOUPE ext. $NEX=2$} and \textit{LOUPE ext. $NEX=3$}) repeat a fixed mask, resulting in accumulated masks with uniform colors, and \textit{Multi-NEX VD} employs a heuristic, non-optimized variable density distribution across both the NEX dimension and \textit{$k$-space}. 
In contrast, NexOP (right column) learns a different distribution strategy: it maintains high-repetition-density sampling in the \textit{$k$-space} center (yellow) while distributing the remaining sampling budget into the periphery (green/blue) to capture high-frequency details.}\label{fig7_masks}
\end{figure*}

Analysis of the learned sampling strategies provides further insights into how NexOP distributes the acquisition budget. The per-repetition sampling rates (Fig.~\ref{fig5:Per_NEX_Sampling}) indicate that the framework allocates the budget in a non-uniform, contrast-dependent manner. Interestingly, in all cases, it arrived at a sampling rate that is \textit{monotonically decreasing} with the repetitions: earlier repetitions receive a larger share of \textit{$k$-space} samples, whereas later repetitions are sampled more sparsely. Furthermore, the learned probability maps (Fig.~\ref{fig6_maps}) show that earlier repetitions generally cover a broader range of spatial frequencies, whereas later repetitions tend to concentrate their sampling density near the \textit{$k$-space} center. A quantitative analysis of the sampling probability distributions (Appendix~\ref{secA_maps_analysis}, Table~\ref{tab:std_summary}) supports this observation: in most experimental settings, the distributions learned by NexOP exhibit decreasing standard deviations from the first to the last repetition, indicating a general trend of narrowing spatial frequency coverage along the NEX dimension. Furthermore, the resulting accumulated masks (Fig.~\ref{fig7_masks}) demonstrate that NexOP learns a \textit{multi-dimensional} variable-density strategy, extending the conventional within-acquisition variable density to the NEX dimension. 

On the reconstruction side, our multi-NEX unrolled architecture, with its Multi-Repetition Data Consistency (MR-DC) blocks, enables reconstructing a single high-fidelity image from multi-NEX data, a task that well-established single-acquisition unrolled networks are generally not designed to perform. Because NexOP learns a monotonically decreasing sampling rate across repetitions, the highly sparse sampling of later repetitions (in some settings below 5\% of \textit{$k$-space}) would be insufficient for independent, per-repetition reconstruction. Our network addresses this by enabling shared image-domain processing of data acquired across different repetitions, whereas data consistency is enforced separately for each repetition. Consequently, the densely sampled first repetition effectively serves as an implicit prior that guides reconstruction from the sparser later ones.

Our framework is modular: the sampling module and the reconstruction module can each be replaced independently. This design facilitates integration with other DL reconstruction methods. Moreover, NexOP optimizes sampling \textit{densities} (probability maps) rather than a single fixed mask, allowing different mask realizations to be drawn per scan and providing additional flexibility in practice.
\purple{In summary, this learned strategy consistently outperforms all competing methods, including both fixed sampling schemes and DL-based approaches that do not optimize across the NEX dimension, demonstrating that joint optimization of the sampling distribution across repetitions is key to achieving high reconstruction quality in the low-SNR regime.}

\bmsubsection{Relation to prior work}

\textbf{DL-based sampling optimization.} Over the recent years, a growing body of work has developed frameworks for jointly optimizing \textit{$k$-space} sampling and image reconstruction. For example, LOUPE \cite{bahadir2020deep} introduced end-to-end learning of a Cartesian undersampling mask jointly with a reconstruction network. J-MoDL \cite{Aggarwal2020} extended this idea to continuous, non-Cartesian sampling locations within a model-based DL framework. PILOT \cite{Weiss2021} incorporated MRI hardware constraints to learn physically feasible non-Cartesian trajectories. Other approaches include SPARKLING \cite{chaithya2022optimizing}, BJORK \cite{wang2022bjork}, BASS \cite{zibetti2021fast}, SUNO \cite{Gautam2025}, and recent work on optimizing sampling patterns using diffusion-based generative models \cite{ravula2023optimizing}. However, these DL-based sampling-optimization frameworks primarily optimize the sampling for a single \textit{$k$-space} acquisition and do not explicitly model the allocation of samples across repeated acquisitions, i.e., across the NEX dimension that is inherent to low-field MRI protocols. NexOP addresses this gap by jointly optimizing \textit{$k$-space} sampling and reconstruction across the multi-NEX domain.

\textbf{NEX-aware acquisition strategies.} Only a limited number of studies have explored how acquisition budgets can be distributed across repetitions. First, Schoormans et al.\ \cite{Schoormans2020} introduced compressed sensing with variable-density averaging, which strategically concentrates signal averages in the low-frequency center of \textit{$k$-space} to improve image quality for low-SNR acquisitions. Second, in our recent work \cite{waddington2022accelerating,shimron2024accelerating}, we demonstrated improved reconstruction quality when samples are distributed across the NEX dimension under a fixed budget, relative to a single fully sampled \textit{$k$-space}. Together, these contributions established that NEX-distributed acquisitions can be useful for the low-SNR regime; nevertheless, both studies relied on pre-defined variable-density masks and classical compressed sensing reconstruction rather than learned sampling or deep learning. NexOP advances beyond these approaches by learning the sampling distribution across the NEX dimension jointly with a tailored DL reconstruction network in an end-to-end fashion. To our knowledge, this combination of NEX-adaptive sampling optimization and multi-repetition DL reconstruction has not been previously demonstrated.

\subsection{Conclusion}
This work presented NexOP, a framework that jointly optimizes the distribution of \textit{$k$-space} samples across repetitions and a DL-based reconstruction network designed for multi-NEX data. Our experiments demonstrated that this joint optimization consistently improves reconstruction quality compared to existing approaches, without increasing the overall acquisition budget. These findings suggest that the NEX dimension represents a largely unexploited degree of freedom in MRI acquisition design, particularly for SNR-limited settings such as low-field and portable systems, where multiple averages are routinely acquired. By enabling shorter scans or higher image quality at fixed scan duration, optimized multi-NEX acquisition strategies have the potential to improve the clinical utility of affordable MRI systems. The modular architecture of NexOP further supports this goal, as the sampling and reconstruction components can be adapted independently to different scanner configurations and imaging protocols. Furthermore, the framework can be readily expanded to additional anatomies, pulse sequences, and low-field MRI systems of various field strengths.

\bmsection*{Author contributions}
Conceptualization: T.O. and E.S.; Coding and performing experiments: T.O.; Manuscript writing and editing: T.O. and E.S.; Supervision: E.S.

\bmsection*{Acknowledgments}
Funded by the European Union (ERC, NEW-CONTRAST-MRI, 101222825). Views and opinions expressed are however those of the author(s) only and do not necessarily reflect those of the European Union or the European Research Council Executive Agency. Neither the European Union nor the granting authority can be held responsible for them.

\vspace{1em }

\bibliography{refs}

\begin{appendix}


\bmsection{Probability maps analysis}\label{secA_maps_analysis}

\begin{figure*}[!t]
\centerline{\includegraphics[width=0.9\textwidth]{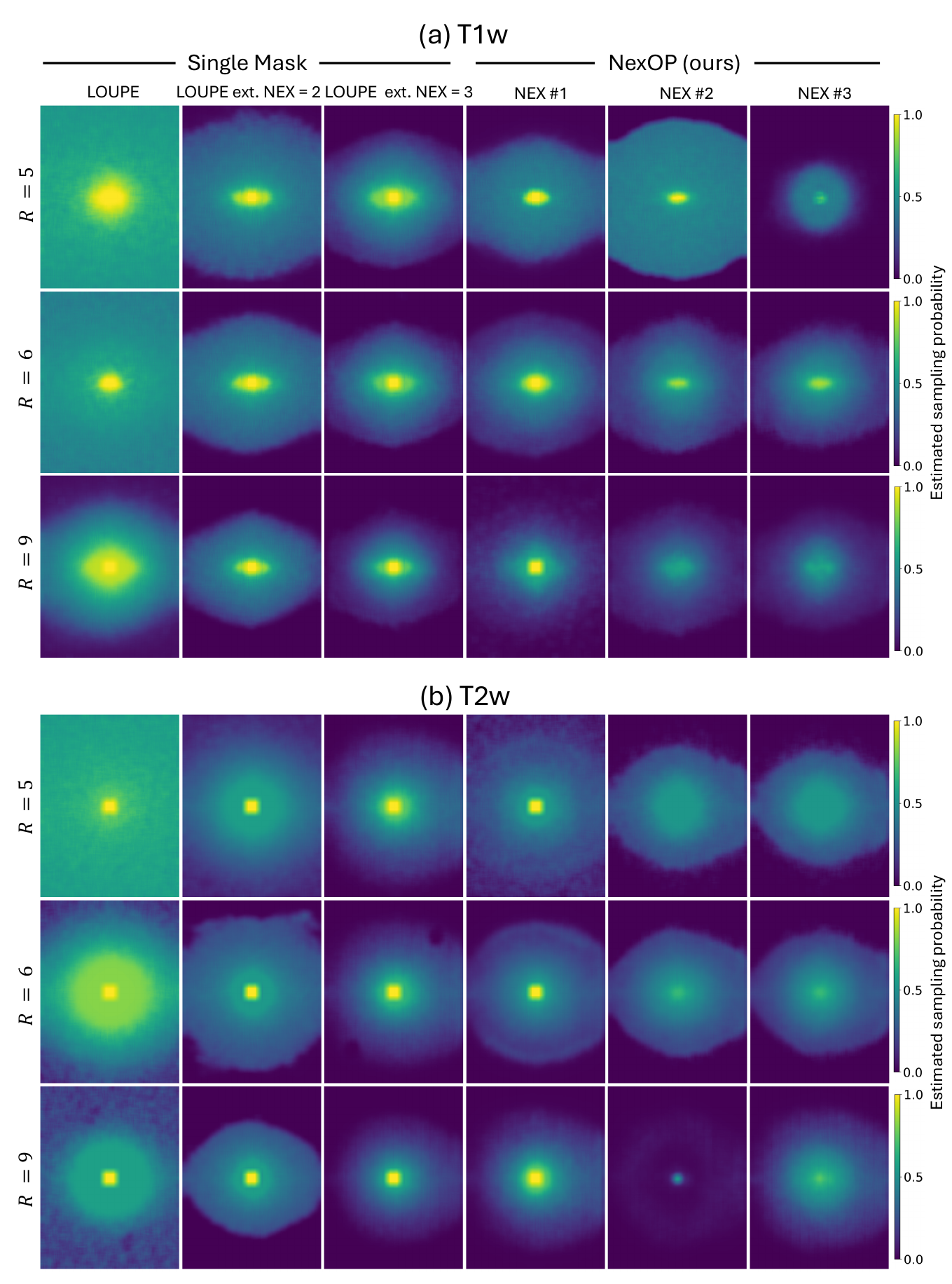}}
\caption{\textbf{Smoothed, numerical probability maps learned by the LOUPE-based and NexOP methods.} The maps were computed by applying a $10 \times 10$ uniform mean filter to the learned probabilities $q$. They are presented for different experimental settings: (a) T1w; and (b) T2w data, for acceleration factors ranging from $R=5$ to $9$.}
\label{fig_maps_analysis}
\end{figure*}

To evaluate the spatial \textit{$k$-space} sampling distribution learned by the different methods, we analyzed the learned probability maps $q$ for the various experimental settings studied here - T1w and T2w contrasts, at acceleration factors of $R = 5,\,6,\,9$. First, to reduce localized noise and visualize smooth sampling density trends, the raw probability maps were smoothed using a 2D uniform mean filter with a $10 \times 10$ pixel kernel. The resulting estimated, spatially smooth probability maps are presented in Fig. \ref{fig_maps_analysis}. As described earlier, each LOUPE method learns a single sampling probability map and repeats it across repetitions, whereas NexOP learns a distinct probability map for each repetition; therefore, a single map is shown for each LOUPE method, whereas three maps are shown for our NexOP method. Because the smoothing procedure is applied uniformly to all maps, the resulting distributions may be slightly broader than the original ones; nevertheless, the standard deviations remain useful for comparing trends across methods and repetitions.

Following the smoothing operation, each map was normalized by its total sum, yielding approximated two-dimensional discrete probability distributions, $\pi(u, v)$, such that $\sum_{u}\sum_{v} \pi(u, v) = 1$. 

To measure the spread of these distributions along the axes of  \textit{$k$-space}, we calculated the marginal standard deviations ($\sigma_u$ and $\sigma_v$) using statistical moments. The first moments (centers of mass) were calculated as:
$$E[u] = \sum_{u}\sum_{v} u \cdot \pi(u, v)$$
$$E[v] = \sum_{u}\sum_{v} v \cdot \pi(u, v)$$

The second moments were then calculated to determine the variance:
$$E[u^2] = \sum_{u}\sum_{v} u^2 \cdot \pi(u, v)$$
$$E[v^2] = \sum_{u}\sum_{v} v^2 \cdot \pi(u, v)$$

Finally, the variances were computed by $\mathrm{Var}(u) = E[u^2] - (E[u])^2$ and $\mathrm{Var}(v) = E[v^2] - (E[v])^2$, allowing us to extract the standard deviations:
$$\sigma_u = \sqrt{\mathrm{Var}(u)}$$
$$\sigma_v = \sqrt{\mathrm{Var}(v)}$$

These standard deviations provide a quantitative measure of how broadly each sampling distribution covers \textit{$k$-space}. A comparison of these standard deviations across the LOUPE and NexOP methods for both T1w and T2w contrasts is summarized in Table \ref{tab:std_summary}. In this setting, larger values reflect broader spatial-frequency coverage, whereas lower values reflect a stronger concentration near the low-frequency center.

Notably, in most settings the distributions learned by NexOP exhibit decreasing standard deviations from the first to the last repetition, with the last repetition generally showing the narrowest distribution; this suggests that later repetitions, to which fewer samples are allocated, tend to concentrate their sampling probability near the \textit{$k$-space} center, whereas earlier repetitions cover a wider range of spatial frequencies. Overall, this analysis provides further quantitative support for the multi-dimensional variable-density strategy described in Section~\ref{sec_discussion}. 

\begin{table}[htbp]
\centering
\caption{Marginal standard deviations ($\sigma_u$, $\sigma_v$) of the two-dimensional smoothed probability distributions derived from the learned probability maps, as described in Section~\ref{secA_maps_analysis}. The standard deviations are provided in units of $k$-space grid points, for various acceleration factors ($R =5,\,6,\,9$) and tissue contrasts (T1w, T2w). Each LOUPE-based method learns a single probability map, which is repeated across the NEX dimension, whereas NexOP learns a distinct map per repetition. \purple{Therefore, three sets of standard deviations are provided for the NexOP method, for each experimental setting.}}
\label{tab:std_summary}
\begin{tabular}{l cccc}
\toprule
 & \multicolumn{2}{c}{\textbf{T1w}} & \multicolumn{2}{c}{\textbf{T2w}} \\
\cmidrule(lr){2-3} \cmidrule(lr){4-5}
 \textbf{Method} & $\sigma_u$ & $\sigma_v$ & $\sigma_u$ & $\sigma_v$ \\

\midrule
\multicolumn{5}{l}{\textbf{R = 5}} \\
  LOUPE ($NEX=1$) & 54.9 & 70.9 & 55.5 & 72.3 \\
  LOUPE-ext ($NEX=2$) & 51.3 & 55.4 & 50.6 & 58.8 \\
  LOUPE-ext ($NEX=3$) & 45.0 & 39.9 & 45.5 & 45.2 \\
\cmidrule{2-5}
  NexOP NEX\#1 & 49.4 & 42.2 & 50.4 & 59.4 \\
  NexOP NEX\#2 & 52.6 & 56.5 & 46.7 & 41.3 \\
  NexOP NEX\#3 & 29.6 & 26.7 & 45.7 & 39.4 \\

\midrule
\multicolumn{5}{l}{\textbf{R = 6}} \\
  LOUPE ($NEX=1$) & 55.0 & 70.5 & 51.7 & 60.3 \\
  LOUPE-ext ($NEX=2$) & 49.9 & 45.1 & 49.4 & 51.8 \\
  LOUPE-ext ($NEX=3$) & 42.9 & 36.2 & 43.7 & 41.4 \\
\cmidrule{2-5}
  NexOP NEX\#1 & 45.9 & 42.6 & 46.0 & 45.7 \\
  NexOP NEX\#2 & 45.1 & 42.2 & 45.9 & 40.5 \\
  NexOP NEX\#3 & 44.0 & 37.6 & 45.0 & 39.1 \\

\midrule
\multicolumn{5}{l}{\textbf{R = 9}} \\
  LOUPE ($NEX=1$) & 49.1 & 49.4 & 51.5 & 61.7 \\
  LOUPE-ext ($NEX=2$) & 45.1 & 34.2 & 44.3 & 36.0 \\
  LOUPE-ext ($NEX=3$) & 37.1 & 31.8 & 39.1 & 35.4 \\
\cmidrule{2-5}
  NexOP NEX\#1 & 43.6 & 41.9 & 42.3 & 39.9 \\
  NexOP NEX\#2 & 43.0 & 39.1 & 51.1 & 49.3 \\
  NexOP NEX\#3 & 41.8 & 36.7 & 43.2 & 41.7 \\

\bottomrule
\end{tabular}
\end{table}


\end{appendix}
\clearpage



\end{document}